\documentclass[twocolumn,secnumarabic,amssymb, nobibnotes, aps, prd, superscriptaddress]{revtex4-1}
\usepackage{amsmath}
\usepackage{graphicx}
\usepackage{mathtools}
\usepackage{epstopdf}

\newcommand{\angstrom}{\mbox{\normalfont\AA}}
\begin{document}

\title{Nonlinear response of biased bilayer graphene at terahertz frequencies}%

\author{Riley McGouran}%
\affiliation{Department of Physics, Engineering Physics and Astronomy, Queen’s University, Kingston, Ontario K7L 3N6, Canada}
\author{Marc M. Dignam}%
\affiliation{Department of Physics, Engineering Physics and Astronomy, Queen’s University, Kingston, Ontario K7L 3N6, Canada}
\date{Dec 29, 2016}%

\begin{abstract}
A density-matrix formalism within the length gauge is developed to calculate the nonlinear response of both doped and undoped biased bilayer graphene (BBLG) at terahertz frequencies. Employing a tight-binding model, we derive an effective two-band Hamiltonian with which we calculate the conduction and valence band dispersion, as well as their respective Bloch states. We then solve for the dynamic equations of the density matrix elements, allowing for the calculation of the intraband and interband current densities and the transmitted and reflected terahertz fields. We find that for undoped BBLG with a gap size of 4 meV, the reflected field exhibits a third harmonic amplitude that is 45\% of the fundamental in the reflected field (0.07\% of the incident field fundamental) for an incident 1 THz single-cycle pulse with a field amplitude of 2.0  kV/cm. We find for doped BBLG, although the dispersion becomes highly nonparabolic as a bias is applied, the third harmonic is a maximum of 8\% of the fundamental in the reflected field (0.56\% of the incident field fundamental) when there is no bias and diminishes with an increase in bias.
\end{abstract}

\maketitle

\section{Introduction}

In recent years there has been intense research focused on both monolayer graphene (MLG), and unbiased bilayer graphene (UBLG). Much of this interest is due to the remarkable transport properties possessed by both of these materials. This is highlighted by very large carrier mobilities and thermal conductivities. In addition, its mechanical properties (stiffness, strength and flexibility) make it ideal as a basis for new composite materials, while its transparency to white light (95\%) makes it ideal for flexible touch screen displays. It even has potential to serve in new energy applications, including batteries\cite{Novo:graphene2005,  Morozov:GiantIntrinsic2008, geim:riseofgraphene2007, geim2009graphene,  allen2009honeycomb, novoselov2011nobel, novoselov2012roadmap, yoo2011ultrathin, wang2012review, kucinskis2013graphene}.

Furthermore, the nonparabolic gapless electron energy bands found in MLG and UBLG lead to an intriguing nonlinear optical and terahertz (THz) response, with theoretical studies predicting that third-harmonic generation (THG) is expected to be observed\cite{al2014high, al2015optimizing, mcgouran2016nonlinear}. Thus, a portion of the interest in MLG and UBLG arises from the fact that it may be used for THz harmonic generation. In a recent paper we found that for undoped, suspended UBLG, a third harmonic is generated that is 30\% of the fundamental in the reflected field, for an incident 1 THz single-cycle pulse with a field amplitude of 1.5  kV/cm \cite{mcgouran2016nonlinear}.

Harmonic generation is only one of the reasons that MLG and UBLG are so appealing to researchers. Perhaps one of the most exciting properties of bilayer graphene is the ability to open a gap in its energy dispersion by applying an external DC bias perpendicular to the layers. We refer to this system as biased bilayer graphene (BBLG). The conduction and valence bands of BBLG do not touch at the Dirac point, and instead exhibit a gap, the size of which is controlled by the bias\cite{mccann2006asymmetry, castro2007biased, neto2009electronic, mccann2013electronic}. The ability to control the size of the gap makes BBLG the first semiconductor with a widely tunable gap, thus making it potentially very important to the modern fields of nanoelectronics and optoelectronics. 

The application of a bias to bilayer graphene results in a difference in the potential energy of the atoms in the top and bottom layers, thereby breaking the inversion symmetry of the lattice. The breaking of inversion symmetry by the bias leads to more than the opening of a band gap. The symmetry breaking also allows for a non-zero \textit{Berry curvature}. It is well known that in the presence of an in-plane electric field, an electron will acquire an \textit{anomalous} velocity in a direction transverse to the field, and with magnitude proportional to the Berry curvature, $\mathbf{\Omega}(\mathbf{k})$, of the band structure\cite{chang1996berry, xiao2010berry}. This provides an anomalous contribution to the \textit{intraband} current density. We shall derive an explicit expression for the Berry curvature of BBLG, which allows for the calculation of the anomalous velocity, as well as \textit{valley} currents - which are localized around the Dirac points within the first Brillouin zone. Additionally, we find that there is also an anomalous \textit{interband} current, which is also a result of inversion symmetry breaking, and has not yet been reported in the literature. 


As in MLG and UBLG, the absorption of optical and THz radiation in BBLG can be characterized by interband and intraband transitions. For an external bias that only induces a gap on the order of a few meV, the low THz photon energy is sufficient to probe transitions between the valence and conduction bands (interband transitions) in undoped BBLG at low temperatures. These interband transitions will be strongly affected by the presence of an external bias as transitions will be greatly suppressed when the gap size is greater than the energy of the incident THz field. Additionally, the intraband current arising from the carriers in the conduction and valence bands will be affected by the distortion of the band dispersions resulting from the bias. For larger biases (corresponding to gap sizes of a few hundred meV), a 'sombrero' feature in the dispersion is manifest\cite{mccann2006asymmetry, castro2007biased, nicol2008optical}. This feature serves to move the minimum of the bands away from the Dirac point. 

Previous theoretical studies of MLG and UBLG have suggested the presence of a strong nonlinearity at optical and THz frequencies\cite{mikhailov2007non, mikhailovnonlin2009, hendry2010coherent, ishikawa2010nonlinear, wright2009strong, ang2010nonlinear}. Experiments have been performed with the intent of observing THG in mono and mullti-layer graphene. It has been observed by using a 45-layer sample, however it has not yet been successfully observed in MLG or UBLG \cite{bowlan2014ultrafast, paul2013high}. In recent theoretical work on MLG and UBLG, it was shown that if the Fermi level is reduced to within only a few meV of the Dirac point, the magnitude of the interband current is comparable to the intraband current, and a strong nonlinearity in the interband current can arise \cite{al2014high, mcgouran2016nonlinear}. Similarly, we expect that the presence of a tunable gap in the band structure of BBLG may lead to a unique interplay between the interband and intraband current densities, and open the possibility to interesting higher order behavior.

In this paper, we present a derivation of a two-band tight-binding model for the intraband and interband dynamics of undoped and doped suspended BBLG in response to a single-cycle pulse at 1 THz.  We use this model to explore the dependency of the nonlinear response on a number of parameters. Specifically, we study the role of the external bias on third and higher harmonic generation. The current densities and the corresponding harmonics are numerically calculated for both undoped and doped BBLG. We find that the ratio of the amplitude of the third harmonic to the fundamental in the reflected field is larger for undoped BBLG with a gap size of 4 meV, than it is for UBLG under identical conditions. Finally, we examine the nonlinear response of doped BBLG with a number of gap sizes. We find that as the size of the gap increases, the third harmonic amplitude decreases; reaching a maximum for a gap size of zero.

The paper is organized as follows. In section \ref{sec:2},  we first present the results of the tight binding model used to obtain the dispersions and eigenvectors corresponding to the low energy conduction and valence bands of BBLG. The eigenvectors allow us to determine expressions for the interband and intraband connection elements, as well as the Berry connections and curvatures of the conduction and valence bands. We then use these expressions to determine the dynamic equations for the density matrix, and the expressions for the intraband and interband current densities. In section \ref{sec:3}, we present the results of numerical simulations for both undoped and doped BBLG. The conclusions are presented in section \ref{sec:4}.

\section{Theory}\label{sec:2}
The calculations that we perform are based on a theoretical approach employing a density-matrix formalism in the length gauge (also known as the electric dipole gauge).  A nearest-neighbor tight-binding model is used to treat the $\pi$-electrons in the graphene, which are taken to provide the conduction electrons\cite{sarma2011electronic}. 
\subsection{Energy Bands}
The tight binding model we employ for BBLG makes use of the solutions found in the case of UBLG. Beginning with the unbiased bilayer Bloch functions \cite{mcgouran2016nonlinear}, we obtain the eigenvalues and eigenvectors of BBLG by solving the characteristic equation for an effective two band Hamiltonian, obtained through the coupling of the lower energy bands: conduction band $c_{1}$ and valence band $v_{2}$, as outlined in Ref. \cite{mcgouran2016nonlinear}. 

The tight-binding expression for the Bloch states is given by 
\begin{equation}\begin{aligned}\psi_{n\mathbf{k}}(\mathbf{r}) & =A_{n}\left(\mathbf{k}\right)\sum_{i}\sum_{\mathbb{\mathbf{R}}}C_{i}^{n}\left(\mathbf{k}\right)\varphi_{pz}\left(\mathbf{r}-\mathbf{R-r}_{i}\right)e^{i\mathbf{k}\cdot\mathbf{R}},\end{aligned}
\end{equation}
where $A_{n}\left(\mathbf{k}\right)$ is a normalization factor, $n$ labels the conduction and valence bands, and the sum is over the Bravais lattice vectors $\mathbf{R}$. The sublattice coefficients, $C_{i}^{n}\left(\mathbf{k}\right)$,
 are associated with the four carbon atoms within the unit cell; the $\varphi_{pz}\left(\mathbf{r}\right)$ are the $2p_{z}$ orbitals of carbon. The index $i$ indicates a sum over the basis vectors $\mathbf{r}_{A_{1}},\,\mathbf{r}_{B_{1}},\,\mathbf{r}_{A_{2}},\,\mathbf{r}_{B_{2}}$, which give the position of sublattice sites $A_{1}$ and $B_{1}$  in the top layer, and $A_{2}$ and $B_{2}$ in the bottom layer. Explicitly, they are given by $\mathbf{r}_{A_{1}}=0$, $\mathbf{r}_{B_{1}}=a_{o}\widehat{\mathbf{x}}$, $\mathbf{r}_{A_{2}}=-a_{o}\widehat{\mathbf{x}}$ and $\mathbf{r}_{B_{2}}=0$.

Formally, when an external bias is applied perpendicularly to the plane of the bilayer, we can express the Hamiltonian of BBLG in the basis of the sublattice Bloch states as
\begin{equation}
\begin{aligned}\mathcal{H}_{b} & =\left(\begin{array}{cccc}
a & f\left(\mathbf{k}\right)t_{\parallel} & 0 & t_{\perp}\\
f\left(\mathbf{k}\right)^{*}t_{\parallel} & a & 0 & 0\\
0 & 0 & -a & f\left(\mathbf{k}\right)t_{\parallel}\\
t_{\perp} & 0 & f\left(\mathbf{k}\right)^{*}t_{\parallel} & -a
\end{array}\right),\end{aligned}\label{eq:ham}
\end{equation}
where, due to the DC bias, the potential energy difference between the atoms in the top and bottom layers is $2a$. The vector of the sublattice coefficients is given by $$\Bigl\langle\mathbf{k}_{n}\Bigr|=\left(C_{A_{1}}^{n*}\left(\mathbf{k}\right),\, C_{B_{1}}^{n*}\left(\mathbf{k}\right),\, C_{A_{2}}^{n*}\left(\mathbf{k}\right),\, C_{B_{2}}^{n*}\left(\mathbf{k}\right)\right).$$ Here $\mathbf{k}$ is the crystal momentum and the function $f\left(\mathbf{k}\right)\equiv(1+e^{-i\mathbf{k}\cdot\mathbf{a}_{1}}+e^{-i\mathbf{k}\cdot\mathbf{a}_{2}})$ is a result of the nearest-neighbor intralayer electron hopping, where the $\mathbf{a}_i$ are the primitive translation vectors of graphene, given explicitly by 
\begin{equation}\begin{aligned}\mathbf{a}_{1} & =\frac{3a_{o}}{2}\widehat{\mathbf{x}}+\frac{\sqrt{3}a_{o}}{2}\widehat{\mathbf{y}},\end{aligned}
\begin{aligned}\mathbf{a}_{2} & =\frac{3a_{o}}{2}\widehat{\mathbf{x}}-\frac{\sqrt{3}a_{o}}{2}\widehat{\mathbf{y}}.\end{aligned}
\end{equation} 
Here $a_{o}$ is the nearest-neighbor separation $(a_{o}\simeq1.42\, \angstrom)$. Also, the \textit{intralayer} hopping energy, $t_{\parallel}$, and the \textit{interlayer} hopping energy, $t_{\perp}$, are approximately equal to 3.03 eV and 0.3 eV, respectively \cite{malard2007probing, zhang2008}. 

We separate the BBLG Hamiltonian into two parts: $\mathcal{H}_{b} =\mathcal{H}_{u}+V,$ where $\mathcal{H}_{u}$ is the Hamiltonian for unbiased bilayer graphene, and $V$ is the matrix representing the potential due to the external bias,
\begin{equation}
\begin{aligned}V & =\left(\begin{array}{cccc}
a & 0 & 0 & 0\\
0 & a & 0 & 0\\
0 & 0 & -a & 0\\
0 & 0 & 0 & -a
\end{array}\right).\end{aligned}
 \end{equation}
Because photons in pulses at THz frequencies possess energies on the order of tens of meV or less, the interband carrier transitions resulting from THz absorption occur almost exclusively between the $c_{1}$ and $v_{2}$ bands (low energy bands). For all biases of interest in this work, the next lowest energy transitions between $v_{2}\rightarrow c_{2}$ and $v_{1}\rightarrow c_{1}$ occur at approximately 75 THz, which is a much higher frequency than we are considering here. Thus, we may obtain an effective biased Hamiltonian, $\mathcal{H}_{b}^{(2)}$, in the basis of the low energy Bloch functions of UBLG (here the superscript (2) denotes that we use as our basis only the Bloch functions that correspond to the conduction and valence bands closest in energy to the Dirac point: $c_{1}$ and $v_{2}$). 


We take our trial variational wavefunction to be a linear combination of the low energy Bloch functions of UBLG:
\begin{equation}
\begin{aligned}\Bigl|\mathbf{k}^{b}\Bigr\rangle & =\sum_{j}a_{j}(\mathbf{k})\Bigl|\mathbf{k}_{j}\Bigr\rangle,\end{aligned}\label{eq:bloch}
\end{equation}
where the $a_{j}(\mathbf{k})$ are expansion coefficients that are determined by solving for the eigenvectors of $\mathcal{H}_{b}^{(2)}$, and the $\Bigl|\mathbf{k}_{j}\Bigr\rangle$ are eigenstates of $\mathcal{H}_{u}$ where $j$ takes on the values $c_{1},\,v_{2}$. Explicitly, we find that in the original sublattice basis, the unbiased eigenstates are given by \cite{mcgouran2016nonlinear}
\begin{equation}
\Bigl|\mathbf{k}_{c_{1}}\Bigr\rangle=\frac{1}{2}\left(\begin{array}{c}
-\left(\frac{\tilde{\epsilon}-t_{\perp}}{\tilde{\epsilon}}\right)^{1/2}\\
-\left(\frac{\tilde{\epsilon}+t_{\perp}}{\tilde{\epsilon}}\right)^{1/2}e^{-i\chi}\\
\left(\frac{\tilde{\epsilon}+t_{\perp}}{\tilde{\epsilon}}\right)^{1/2}e^{i\chi}\\
\left(\frac{\tilde{\epsilon}-t_{\perp}}{\tilde{\epsilon}}\right)^{1/2}
\end{array}\right),\label{eq:5}
\end{equation} 

\begin{equation}
\Bigl|\mathbf{k}_{v_{2}}\Bigr\rangle=\frac{1}{2}\left(\begin{array}{c}
\left(\frac{\tilde{\epsilon}-t_{\perp}}{\tilde{\epsilon}}\right)^{1/2}\\
-\left(\frac{\tilde{\epsilon}+t_{\perp}}{\tilde{\epsilon}}\right)^{1/2}e^{-i\chi}\\
-\left(\frac{\tilde{\epsilon}+t_{\perp}}{\tilde{\epsilon}}\right)^{1/2}e^{i\chi}\\
\left(\frac{\tilde{\epsilon}-t_{\perp}}{\tilde{\epsilon}}\right)^{1/2}
\end{array}\right),\label{eq:6}
\end{equation}
where $\tilde{\epsilon}(\mathbf{k})=\sqrt{t_{\perp}^{2}+4|f\left(\mathbf{k}\right)|^{2}}$ and $ e^{i\chi\left(\mathbf{k}\right)}=f\left(\mathbf{k}\right)/\left\vert f\left(\mathbf{k}\right)\right\vert$. In Eqs. (\ref{eq:5}) and (\ref{eq:6}), we have suppressed the explicit $\mathbf{k}$-dependencies for simplicity. 

The matrix elements of $\mathcal{H}_{b}^{(2)}$ are then given by $\mathcal{H}_{b}^{{(2)}^{ij}}=\left\langle \mathbf{k}_{i}\right|\left[\mathcal{H}_{u}+V\right]\left|\mathbf{k}_{j}\right\rangle,$ for $i,j=\{c_{1},v_{2}\}$. Since the unbiased Hamiltonian $\mathcal{H}_{u}$ is diagonal in the basis of the Bloch functions $\left|\mathbf{k}_{j}\right\rangle$, the off-diagonal elements of $\mathcal{H}_{b}^{(2)}$ will be due entirely to the potential $V$. Thus, our effective biased Hamiltonian takes the form
\begin{equation}
\begin{aligned}\mathcal{H}_{b}^{(2)} & =\left[\left(\begin{array}{cc}
\mathcal{H}_{u}^{c_{1}c_{1}} & 0\\
0 & \mathcal{H}_{u}^{v_{2}v_{2}}
\end{array}\right)+\left(\begin{array}{cc}
V^{c_{1}c_{1}} & V^{c_{1}v_{2}}\\
V^{v_{2}c_{1}} & V^{v_{2}v_{2}}
\end{array}\right)\right]\\
\\
 & =\left(\begin{array}{cc}
E_{0}(\mathbf{k}) & \frac{at_{\perp}}{\tilde{\epsilon}(\mathbf{k})}\\
\frac{at_{\perp}}{\tilde{\epsilon}(\mathbf{k})} & -E_{0}(\mathbf{k})
\end{array}\right),
\end{aligned}\label{eq:hameff}
 \end{equation}
where we have that $V^{c_{1}v_{2}}=V^{v_{2}c_{1}}=\frac{at_{\perp}}{\tilde{\epsilon}(\mathbf{k})}$. Here $E_{0}(\mathbf{k})$ is the energy of the low energy conduction band ($c_{1}$) in the unbiased case: $E_{0}(\mathbf{k})=\frac{\tilde{\epsilon}(\mathbf{k})-t_{\perp}}{2}$ \cite{mcgouran2016nonlinear}.  

\begin{figure}[h]
\centering
\includegraphics[width=0.5\textwidth]{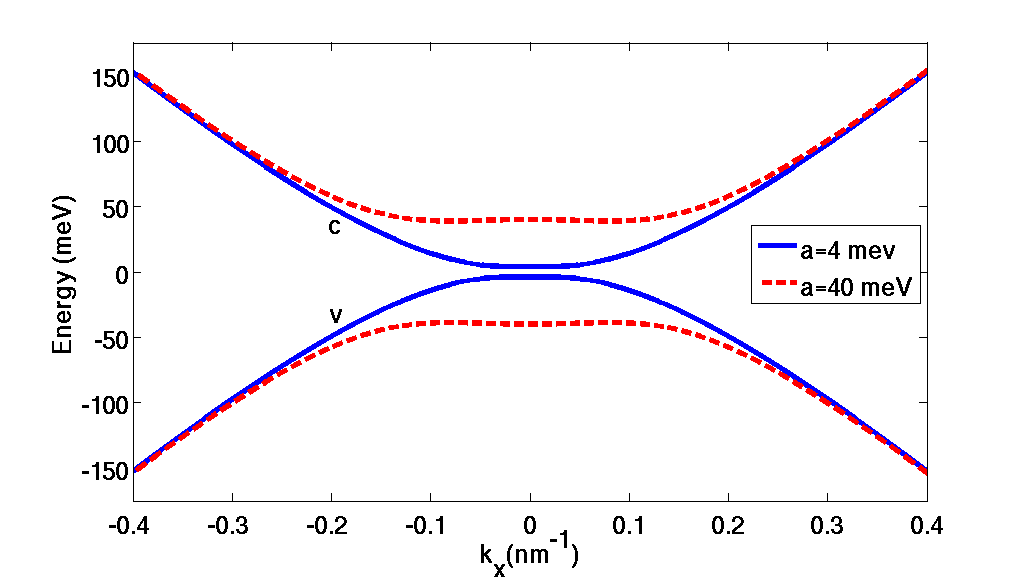}
\caption{Energy bands of BBLG as a function of the crystal momentum $\mathbf{k}$ for two different biases. A gap between conduction and valence bands is present for both bias values (measured in meV); we see a flattening of the bands at the 40 meV bias.\label{fig:disper}}
\end{figure}

We may now solve for the eigenvalues and eigenvectors of $\mathcal{H}_{b}^{(2)}$. Doing so, we have for the dispersions of the biased conduction and valence bands
\begin{equation}
\begin{aligned}E_{c_{1}}^{b}(\mathbf{k}) & =E_{b}(\mathbf{k}),\\
E_{v_{2}}^{b}(\mathbf{k}) & =-E_{b}(\mathbf{k}),\\
E_{b}(\mathbf{k}) & =\sqrt{E_{0}^{2}(\mathbf{k})+S^{2}(\mathbf{k})},
\end{aligned}\label{eq:disper}
\end{equation}
where we have defined $S(\mathbf{k})\equiv \frac{at_{\perp}}{\tilde{\epsilon}(\mathbf{k})}.$
The calculated dispersions are shown in Fig. \ref{fig:disper} for two different bias values. As has been found by previous researchers\cite{mccann2006asymmetry, neto2009electronic}, there is an opening of a band gap in the presence of a non-zero bias. The 'sombrero' feature also becomes clearly present as the external bias is increased to larger values.

The corresponding eigenvectors for the biased conduction and valence bands are found by solving for the coefficients of the unbiased Bloch functions, $a_{j}(\mathbf{k}),$ in Eq.(\ref{eq:bloch}). Solving for these coefficients allows us to express the eigenvectors explicitly as
\begin{equation}
\Bigl|\mathbf{k}_{c_{1}}^{b}\Bigr\rangle=A(\mathbf{k})\left(S(\mathbf{k})\Bigl|\mathbf{k}_{c_{1}}\Bigr\rangle-(E_{0}(\mathbf{k})-E_{b}(\mathbf{k}))\Bigl|\mathbf{k}_{v_{2}}\Bigr\rangle\right)\label{eq:con},
\end{equation}
\begin{equation}
\Bigl|\mathbf{k}_{v_{2}}^{b}\Bigr\rangle=A(\mathbf{k})\left((E_{0}(\mathbf{k})-E_{b}(\mathbf{k}))\Bigl|\mathbf{k}_{c_{1}}\Bigr\rangle+S(\mathbf{k})\Bigl|\mathbf{k}_{v_{2}}\Bigr\rangle\right)\label{eq:val},
\end{equation}
where $A$ is a normalization constant given by
\begin{equation}
A(\mathbf{k})=\frac{1}{\sqrt{2E_{b}(\mathbf{k})(E_{b}(\mathbf{k})-E_{0}(\mathbf{k}))}}.
\end{equation}
It is easy to show that these eigenvectors are orthonormal: $\left\langle \mathbf{k}_{i}^{b}\right|\left.\mathbf{k}_{j}^{b}\right\rangle =\delta_{ij}$. We can also show that in the limit $a\rightarrow0$, these expressions for the Bloch functions of the biased conduction and valence bands reduce to the corresponding unbiased Bloch functions, $\Bigl|\mathbf{k}_{c_{1}}\Bigr\rangle\,\&\,\Bigl|\mathbf{k}_{v_{2}}\Bigr\rangle,$ respectively.

As was done in our study on unbiased bilayer graphene, we employ the length gauge in order to model the interaction of BBLG with THz radiation. This method avoids low-frequency divergences that arise when using the velocity gauge \cite{virk2007semiconductor, moss1990band}. The Hamiltonian in the length gauge is expressed as $H=H_{0}-e\mathbf{r}\cdot\mathbf{E}$, where $H_{0}$ is the effective Hamiltonian for BBLG (Eq. \ref{eq:hameff}), $e = -|e|$ is the charge of an electron, $\mathbf{r}$ is the electron position vector, and $\mathbf{E}(t)$ is the THz electric field at the graphene. For normally incident plane waves, the field is taken to be uniform over the graphene sheets.

\subsection{Connection Elements}
Modelling the carrier dynamics in BBLG requires the calculation of the connection elements between the different electron bands. These arise from the matrix elements of the position operator $\mathbf{r}$, between the Bloch states of BBLG \cite{al2014high, aversa1995nonlinear}:

\begin{equation}
\left\langle n,\mathbf{k}\right.\left|\mathbf{r}\right.\left|m,\mathbf{k}^{\prime}\right\rangle =\delta(\mathbf{k}-\mathbf{k}^{\prime})\mathbf{\xi}_{nm}(\mathbf{k})+i\delta_{nm}\mathbf{\nabla_{k}}\delta(\mathbf{k}-\mathbf{k}^{\prime}),
\label{eq:connection}
\end{equation} where the connection elements are defined as

\begin{equation}
\mathbf{\xi}_{nm}(\mathbf{k})=\frac{\left(2\pi\right)^{2}i}{\Omega}\int_{\Omega}d^{3}\mathbf{r}u_{n,\mathbf{k}}^{\ast}\left(\mathbf{r}\right)\mathbf{\nabla}_{\mathbf{k}}u_{m,\mathbf{k}}\left(\mathbf{r}\right).
\end{equation} 
Here $\Omega$ is the volume of a unit cell and $u_{n\mathbf{k}}(\mathbf{r})$ is the periodic part of the Bloch function. We can evaluate this expression using the biased Bloch functions given in Eqs. (\ref{eq:con}) and (\ref{eq:val}). In these calculations we ignore the overlap of the wave functions on different atomic sites. To simplify notation, in all that follows we shall replace $c_{1}$ with $c$ and $v_{2}$ with $v$ and shall simply refer to them as conduction and valence bands.

Due to the symmetry between the sublattices, the conduction and valence states in graphene are degenerate at two \textit{Dirac points}, given by: \begin{equation}\begin{aligned}\mathbf{K}a_{o} & =\frac{4\pi}{3\sqrt{3}}\widehat{\mathbf{y}},\\
\mathbf{K}^{\prime}a_{o} & =\frac{8\pi}{3\sqrt{3}}\widehat{\mathbf{y}}.
\end{aligned}\end{equation}
For energies close to the Dirac points - within a few hundred meV - we can expand the crystal momentum around the Dirac points as $\mathbf{k}=\mathbf{K}+\delta\mathbf{k}$ and $\mathbf{k}^{\prime}=\mathbf{K}^{\prime}+\delta\mathbf{k}$, where $\delta\mathbf{k}=k_{x}\hat{\mathbf{x}}+k_{y}\hat{\mathbf{y}}$. With this expansion, we find that the biased \textit{interband} connection element between the conduction and valence bands, $\mathbf{\xi}_{vc}^{b}(\mathbf{k}),$ is given by
\begin{equation}
\begin{aligned}\mathbf{\xi}_{vc}^{b}\left(\mathbf{K}+\delta\mathbf{k}\right) & =\frac{E_{0}}{E_{b}}\mathbf{\xi}_{vc}(\mathbf{k})-\frac{iS}{4E_{b}^{2}}\left(\tilde{\epsilon}+2E_{0}\right)\frac{\alpha k}{\tilde{\epsilon}^2}\hat{\mathbf{k}},\end{aligned}\label{eq:connec}
\end{equation}
where for simplicity, the explicit $\mathbf{k}$-dependencies of $E_{b}(\mathbf{k}),\,E_{0}(\mathbf{k})$, $S(\mathbf{k})$ and $\tilde{\epsilon}(\mathbf{k})$ have been suppressed, and \textit{k} is the magnitude of the crystal momentum $k\equiv|\delta\mathbf{k}|$. Here we have defined the constant $\alpha=4\hbar^{2}v_{F}^2$, where $v_{F}=3a_{0}t_{\perp}/2\hbar$ is the Fermi velocity. We have also used the results of our calculations for the unbiased interband and intraband connection elements: $i\Bigl\langle\mathbf{k}_{v}\Bigr|\nabla_{k}\Bigl|\mathbf{k}_{c}\Bigr\rangle=\mathbf{\xi}_{vc}(\mathbf{k})$, and $i\Bigl\langle\mathbf{k}_{n}\Bigr|\nabla_{k}\Bigl|\mathbf{k}_{n}\Bigr\rangle=\mathbf{\xi}_{nn}(\mathbf{k})=0$, respectively\cite{mcgouran2016nonlinear}. Around the $\mathbf{K}$-Dirac point we can express $\mathbf{\xi}_{vc}(\mathbf{k})$ as
\begin{equation}
\mathbf{\xi}_{vc}\left(\mathbf{K}+\delta\mathbf{k}\right)=\left(\frac{\tilde{\epsilon}+t_{\perp}}{2\tilde{\epsilon}}\right)\frac{\widehat{\mathbf{\theta}}}{k}.
\end{equation}

In these expressions for the connection elements, $\delta\hat{\mathbf{k}}=cos(\theta)\hat{\mathbf{x}}+sin(\theta)\hat{\mathbf{y}}$ and $\hat{\mathbf{\theta}}=-sin(\theta)\hat{\mathbf{x}}+cos(\theta)\hat{\mathbf{y}}$ are, respectively, the radial and angular unit vectors in cylindrical coordinates with the origin at the $\mathbf{K}$-Dirac point. In comparison to that for UBLG, we find that the biased connection element has both $\hat{\mathbf{\theta}}$ and $\hat{\mathbf{k}}$ components (Eq. (\ref{eq:connec})). It is easy to show that in the limit $a\rightarrow0$, we have that $\mathbf{\xi}_{vc}^{b}(\mathbf{k})\rightarrow\mathbf{\xi}_{vc}(\mathbf{k}),$ as expected. We shall see that the two component nature of the biased connection element leads to a significant interband current density contribution that is absent in the unbiased case.

Next, we calculate the biased \textit{intraband} connection elements. As discussed previously, these intraband connection elements are identical to Berry connections\cite{chang1996berry}\cite{xiao2010berry}, and were shown to be zero in the case of UBLG \cite{mcgouran2016nonlinear}. However, in the presence of a non-zero bias, the intraband connections do not vanish. They are given explicitly by 
\begin{equation}
\begin{aligned}\mathbf{\xi}_{cc}^{b}(\mathbf{k}) & =\frac{S}{E_{b}}\mathbf{\xi}_{vc}(\mathbf{k}),\\
\mathbf{\xi}_{vv}^{b}(\mathbf{k}) & =-\mathbf{\xi}_{cc}^{b}(\mathbf{k}).
\end{aligned}
\end{equation}

To calculate the full nonlinear response of BBLG, we also require the expressions for the biased interband and intraband connection elements around the $\mathbf{K}^{\prime}$-Dirac point. Explicitly, these are given by
\begin{equation}
\begin{aligned}\mathbf{\xi}_{vc}^{b}\left(\mathbf{K}^{\prime}+\delta\mathbf{k}\right) & =-\frac{E_{0}}{E_{b}}\mathbf{\xi}_{vc}(\mathbf{k})-\frac{iS}{4E_{b}^{2}}\left(\tilde{\epsilon}+2E_{0}\right)\frac{\alpha k}{\tilde{\epsilon}^2}\hat{\mathbf{k}},\end{aligned}\label{eq:connecp}
\end{equation}
and
\begin{equation}
\begin{aligned}\mathbf{\xi}_{cc}^{b}\left(\mathbf{K}^{\prime}+\delta\mathbf{k}\right) & =-\mathbf{\xi}_{cc}^{b}\left(\mathbf{K}+\delta\mathbf{k}\right),\\
\mathbf{\xi}_{vv}^{b}\left(\mathbf{K}^{\prime}+\delta\mathbf{k}\right) & =-\mathbf{\xi}_{vv}^{b}\left(\mathbf{K}+\delta\mathbf{k}\right),
\end{aligned}
\end{equation}
respectively. Thus, we find that the $\hat{\mathbf{\theta}}$-component of the interband connection element, $\mathbf{\xi}_{vc}^{b}(\mathbf{k})$, changes sign as we move from $\mathbf{K}\rightarrow \mathbf{K}^{\prime}$, but the $\hat{\mathbf{k}}$-component does not. We also find that both of the intraband connection elements change sign upon moving from $\mathbf{K}\rightarrow \mathbf{K}^{\prime}$, as they only have components in the $\hat{\mathbf{\theta}}$ direction.

\subsection{Berry Curvature}
\begin{figure}[h]
\centering
\includegraphics[width=0.5\textwidth]{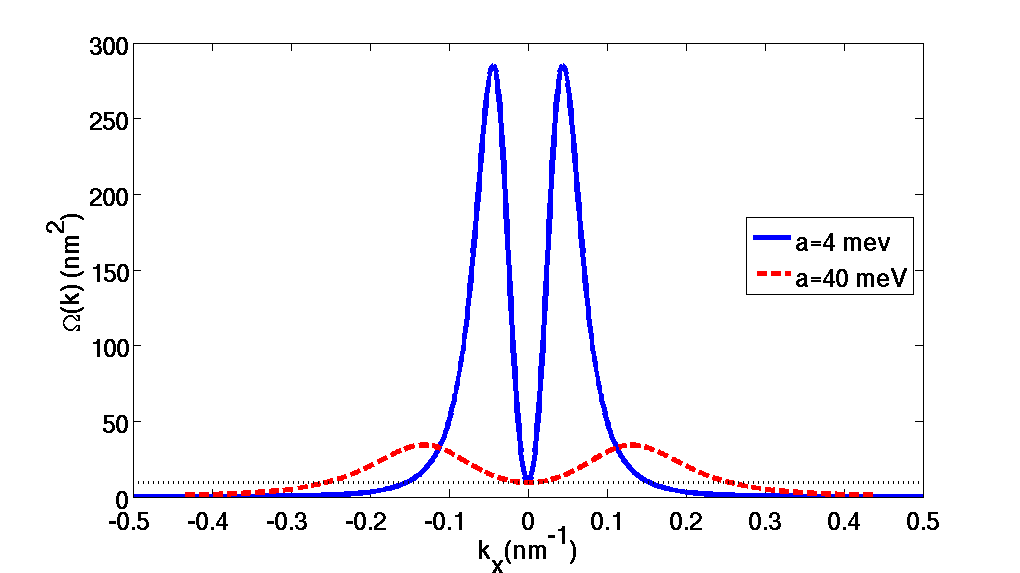}
\caption{Absolute value of the Berry curvature of the conduction band in BBLG near the Dirac point. Results are given by Eq. (\ref{eq:berry}), shown for biases of 4 and 40 meV. Each curvature has the value $2\hbar^{2}v_{F}^{2}/t_{\perp}^{2}$
  at the Dirac point $(\hbar\mathbf{k}=0)$, shown by the dotted black line.\label{fig:Berry}}
\end{figure}

The non-zero intraband connection elements lead directly to non-zero Berry curvatures of the respective bands. This is one of the factors that makes BBLG an interesting system to study. To calculate the Berry curvature of the conduction and valence bands, we simply need to take the curl of the Berry connections of these bands: $\begin{aligned}\mathbf{\Omega}_{nn}(\mathbf{k}) & =\nabla_{\mathbf{k}}\times\mathbf{\xi}_{nn}^{b}(\mathbf{k}).\end{aligned}$ Explicitly, for the conduction band, we obtain for the Berry curvature around the $\mathbf{K}$-point,
\begin{equation}
\begin{aligned}\mathbf{\Omega}_{cc}(\mathbf{k}) & =\nabla_{\mathbf{k}}\times\frac{S}{E_{b}}\mathbf{\xi}_{vc}(\mathbf{k}),\\
 & =\frac{-\alpha S}{4E_{b}^{3}\tilde{\epsilon}^{3}}\left(2t_{\perp}E_{b}^{2}+E_{0}(2\tilde{\epsilon}-t_{\perp})(\tilde{\epsilon}+t_{\perp})\right)\hat{\mathbf{z}}.
\end{aligned}\label{eq:berry}
\end{equation}
From the relationship we have between the Berry connections of the conduction and valence bands, we see that $\mathbf{\Omega}_{cc}(\mathbf{k})=-\mathbf{\Omega}_{vv}(\mathbf{k}).$ As a check, we can determine the Berry curvature in the limit of very small electron momentum $\mathbf{k}$. One can show in this limit that Eq. (\ref{eq:berry}) reduces to 
\begin{equation}
\begin{aligned}\lim_{k\rightarrow0}\mathbf{\Omega}_{cc}(\mathbf{k}) & =-\left[\frac{2\gamma\Delta k^{2}}{\left(k^{4}+\gamma^{2}\Delta^{2}\right)^{3/2}}+\frac{2}{\gamma^{2}}\right]\hat{\mathbf{z}},\end{aligned}
\end{equation}
where $\gamma=t_{\perp}/\hbar v_{F}$, and $\Delta=a/\hbar v_{F}$. The above approximation predicts a non-zero Berry curvature at the Dirac point, i.e., when $\hbar\mathbf{k}$ is zero. We find explicitly that it reduces to the value $2\hbar^{2}v_{F}^{2}/t_{\perp}^{2}$ at the Dirac point, irrespective of the strength of the bias.  We can see this  in Fig. \ref{fig:Berry} where we plot Eq. (\ref{eq:berry}) vs. the electron momentum for two different bias values. This differs from other calculations of the Berry curvature found in the literature, where the curvature is predicted to go to zero at the Dirac point\cite{zhang2013valley}. Note that there is no contradiction in this because when $a\rightarrow0$, the bands touch and so there is an ambiguity as to what the (degenerate) states are at $k=0$. We will see later that the Berry curvature contributes a first order, \textit{anomalous} contribution to the intraband current density around each Dirac point, as mentioned in the introduction.
\subsection{Current Density}
 In order to calculate the current density in BBLG, we require the dynamic equations for the reduced density matrix elements. The dynamic equations for BBLG take the same form as those of UBLG \cite{mcgouran2016nonlinear} and are given by 
\begin{equation}
\begin{alignedat}{1}\frac{d\rho_{nm}(\mathbf{k})}{dt} & =i\frac{e}{\hbar}\mathbf{E}(t)\cdot\sum_{l}(\mathbf{\xi}_{nl}^{b}(\mathbf{k})\rho_{lm}(\mathbf{k})-\rho_{nl}(\mathbf{k})\mathbf{\xi}_{lm}^{b}(\mathbf{k}))\\
 & -\frac{e}{\hbar}\mathbf{E}(t)\cdot\left[\nabla_{\mathbf{k}}\rho_{nm}(\mathbf{k})-i\rho_{nm}(\mathbf{k})(\mathbf{\xi}_{nn}^{b}(\mathbf{k})-\mathbf{\xi}_{mm}^{b}(\mathbf{k}))\right]\\
 & -i\omega_{nm}(\mathbf{k})\rho_{nm}(\mathbf{k})\\
 & -\frac{\rho_{nm}\left(\mathbf{k}\right)-\delta_{nm}\rho_{nm}^{eq}(\mathbf{k})}{\tau_{nm}},
\end{alignedat}
\end{equation} 
where $n,m=\{c,v\}$, $ \omega_{nm}(\mathbf{k})=[E_{n}^{b}(\mathbf{k})-E_{m}^{b}(\mathbf{k})]/\hbar,$ and $\rho_{nm}^{eq}(\mathbf{k})=f_{n}(\mathbf{k},T)\delta_{nm}$ is the carrier population in equilibrium when $n=m$, and is zero otherwise; $f_{n}(\mathbf{k},T)$ is the Fermi-Dirac distribution with a temperature T. In our numerical work, we model the populations of valence band vacancies ($\rho_{hh}(\mathbf{k})$) rather than valence band electrons ($\rho_{vv}(\mathbf{k})$), as this allows us to only include states near the Dirac point, which greatly reduces computation time. The relationship between the two is simply, $\rho_{hh}(\mathbf{k})=1-\rho_{vv}(\mathbf{k})$.

Because scattering times in graphene are on the order of tens of femtoseconds \cite{bowlan2014ultrafast}\cite{paul2013high}, to accurately model the THz response we must take into account scattering processes. Following the approach used for MLG and UBLG\cite{al2014high, mcgouran2016nonlinear}, we treat scattering phenomenologically. We introduce an interband decoherence time, $\tau_{nm}$, for the interband coherences, $\rho_{nm}(\mathbf{k})$, where $n\neq m$. We assume the decoherence time to be independent of $\mathbf{k}$. The populations, $\rho_{nn}(\mathbf{k})$, we take to relax back to Fermi-Dirac thermal distributions, $f_{n}(\mathbf{k},T)$, with relaxation times, $\tau_{n}$. As the simulation proceeds, we adjust the temperature of the Fermi-Dirac distribution so that the carriers relax to the correct total carrier populations at each time-step. We neglect interband relaxation since the time taken for the conduction band electrons to relax to the valence band is much longer than intraband scattering times\cite{tielrooij2013}.

In our simulations we use a direct computational approach to solve the above equations. To do so we put $\mathbf{k}$ on a grid and step through time using a Runge-Kutta algorithm. To facilitate this, we make use of balanced difference approximations to the gradients. Given the geometry of the graphene lattice and Brillouin zone, we employ a hexagonal grid with a uniform point density in k space.

The expression for the current density in BBLG can now be determined using the dynamic equations for the density matrix elements. Following the formalism of Aversa and Sipe\cite{aversa1995nonlinear} the current density can be expressed as 
\begin{equation}\mathbf{J}(t)=\frac{e}{m}\text{Tr}\left\{ \mathbf{p}\rho(t)\right\}.\label{eq:Jt}\end{equation} Using the fact that $\frac{\mathbf{p}}{m}=\frac{1}{i\hbar}\left[\mathbf{r},H\right],$ and decomposing the position operator into intraband and interband parts, $\mathbf{r}=\mathbf{r}_{i}+\mathbf{r}_{e},$ we can write this as\cite{aversa1995nonlinear}
\begin{equation}\begin{aligned}\mathbf{J}(t) & =\frac{e}{i\hbar}\text{Tr}\left\{ \left[\mathbf{r},H\right]\rho(t)\right\} \\
 & =\frac{e}{i\hbar}\sum_{n}\sum_{\mathbf{k}}\left\langle n,\mathbf{k}\right.|\left[\mathbf{r}_{i},H\right]\rho(t)\left|n,\mathbf{k}\right\rangle \\
 & +\frac{e}{i\hbar}\sum_{n}\sum_{\mathbf{k}}\left\langle n,\mathbf{k}\right.|\left[\mathbf{r}_{e},H\right]\rho(t)\left|n,\mathbf{k}\right\rangle, 
\end{aligned}\end{equation} where the trace is over the single electron states, and $\rho(t)$ is the reduced density matrix with matrix elements $\rho_{nm}(\mathbf{k})$. The decomposition of the position operator allows us to define the total current density as the sum of an intraband contribution, $\mathbf{J}_{i}$, and an interband contribution, $\mathbf{J}_{e}$. Using our effective Hamiltonian, as well as the matrix elements of the position operator (Eq. \ref{eq:connection}), one may determine expressions for these contributions. The procedure is similar to that presented in recent work on MLG and UBLG\cite{al2015nonperturbative, mcgouran2016nonlinear}. After considerable work, the intraband current density near the Dirac point can be shown to be given by

\begin{equation}\begin{aligned}\mathbf{J_{i}} & =\frac{e}{\hbar}\sum_{\mathbf{k}}\left[\nabla_{\mathbf{\mathbf{k}}}E_{c_{1}}^{b}(\mathbf{k})-e\mathbf{E}\times\Omega_{c_{1}c_{1}}(\mathbf{k})\right]\left(\rho_{cc}(\mathbf{k)}+\rho_{hh}(\mathbf{k)}\right)\\
 & -\frac{2e^{2}}{\hbar}\sum_{\mathbf{k}}\text{Re}\left\{ \rho_{cv}(\mathbf{k})\right\} \mathbf{E}\times\mathbf{\Lambda}(\mathbf{k})\\
 & -\frac{2e^{2}}{\hbar}\sum_{\mathbf{k}}\left[\text{Re}\left\{ \rho_{cv}(\mathbf{k})\left(\mathbf{E}\cdot\mathbf{\nabla_{\mathbf{k}}}\right)\mathbf{\xi}_{vc}^{b}(\mathbf{k})\right\} \right.\\
 & -\left.2\textrm{Im}\left\{ \rho_{cv}(\mathbf{k})\left(\mathbf{E}\cdot\mathbf{\xi}_{cc}^{b}(\mathbf{k})\right)\mathbf{\xi}_{vc}^{b}(\mathbf{k})\right\} \right],
\end{aligned}
\label{eq:3}\end{equation} where we have defined
\begin{equation}
\mathbf{\Lambda}(\mathbf{k})\equiv\left(\nabla_{\mathbf{\mathbf{k}}}-2\mathbf{\xi}_{cc}^{b}(\mathbf{k})\right)\times\mathbf{\xi}_{vc}^{b}(\mathbf{k}).
\end{equation}

Next, we calculate the interband current density by taking the time derivative of the polarization density, 
\begin{equation}
\mathbf{J}_{e}=\frac{d\mathbf{P}_{e}}{dt},
\end{equation}
where the polarization density is given by
\begin{equation}
\begin{aligned}\mathbf{P}_{e} & =\end{aligned}
2e\sum_{\mathbf{k}}\textrm{Re}\left\{ \mathbf{\xi}_{vc}^{b}(\mathbf{k})\rho_{cv}(\mathbf{k})\right\}.\label{eq:4}
\end{equation} 
This procedure is different than what is proposed in Eq. (\ref{eq:Jt}), and ultimately allows for the simplest calculation of the interband current density. 

The sums over $\mathbf{k}$ in the expressions for the intraband and interband current densities are restricted to a region near the $\mathbf{K}$-Dirac point. We also need to take into account the current density near the $\mathbf{K}^{\prime}$ point. For MLG and UBLG, we found that due to the symmetry of the Brillouin zone, the current densities around both Dirac points are identical\cite{al2015nonperturbative, mcgouran2016nonlinear}. To obtain the total current density in these cases, we simply multiplied the results calculated at the $\mathbf{K}$-point by two in order to account for this degeneracy. However, in the case of BBLG, due to the nature of the connection elements and density matrix elements, it is not immediately obvious whether the current densities around each unique Dirac point are identical or not. To deal with this, in our simulation, we calculate the current densities around each individual Dirac point and combine the contributions.

We consider a suspended graphene sample such as employed in experiments on MLG\cite{paul2013high}, and use the time-dependent current densities to calculate the transmitted and the reflected THz fields, using a procedure identical to that used for MLG\cite{al2015nonperturbative}. We have verified convergence in the nonlinear regime by changing the grid density, the extent of the grid, the time-step tolerance, and the polarization of the incident field.

\subsection{Linear Response}
Before considering the nonlinear response of BBLG, we first examine the linear response to an incident THz field. In order to calculate the linear response we need to calculate expressions for the first order density matrix elements. Once we have these, we may then use Eq. (\ref{eq:3}) to express the first order \textit{intraband} current density as
\begin{equation}
\begin{aligned} \mathbf{J_{i}}^{(1)} & =\frac{2e}{\hbar}\sum_{\mathbf{k}}\left\{ \frac{\left(E_{0}-\frac{2S^{2}}{\tilde{\epsilon}}\right)}{E_{b}}\frac{\alpha\mathbf{k}}{2\tilde{\epsilon}}\sum_{\prime}\left(\rho_{cc}^{(1)}(\mathbf{k})+\rho_{hh}^{(1)}(\mathbf{k})\right)\right.\\
 & +\left.e\mathbf{E}\times\mathbf{\Omega}_{cc}(\mathbf{k})\left(\Delta\rho_{cc}^{(0)}(\mathbf{k})+\Delta\rho_{hh}^{(0)}(\mathbf{k})\right)\right\}. 
\end{aligned}\label{eq:intra1}
\end{equation}
Here we have included the factor of 2 to account for spin degeneracy. We have also defined the difference between the zeroth order populations near the $\mathbf{K}$ and $\mathbf{K}^{\prime}$-points as
\begin{equation}
\Delta\rho_{nn}^{(0)}(\mathbf{k})=\rho_{nn}^{(0)}(\mathbf{\mathbf{K+}\mathbf{\delta k})}-\rho_{nn}^{(0)}(\mathbf{\mathbf{K^{\prime}+}\mathbf{\delta k})}\label{eq:diff},
\end{equation}
and the sum of the first order populations near the $\mathbf{K}$ and $\mathbf{K}^{\prime}$-points as
\begin{equation}
\sum_{\prime}\rho_{nn}^{(1)}(
\mathbf{k})=\rho_{nn}^{(1)}(\mathbf{\mathbf{K+}\mathbf{\delta k})}+\rho_{nn}^{(0)}(\mathbf{\mathbf{K^{\prime}+}\mathbf{\delta k})}\label{eq:sum}.
\end{equation}
It is well known that if the band structure of a crystalline solid has a non-zero Berry curvature, the electrons in those bands will acquire a component to their velocity that is transverse to an applied electric field \cite{xiao2010berry}. This component is commonly known as the \textit{anomalous} velocity. We can see in Eq. (\ref{eq:intra1}) that there is a component of the intraband current density that is perpendicular to the direction of the field, the magnitude of which is proportional to the Berry curvature of the conduction band, $\mathbf{\Omega}_{cc}(\mathbf{k})$. This contribution arises from the anomalous velocity of the electrons, and is equivalent to a Hall current. We see that there is a first order contribution to the anomalous intraband current density around each individual Dirac point, however, the full anomalous contribution goes to zero because the zeroth order population difference, given by Eq.(\ref{eq:diff}), vanishes. In this situation, the sum of the populations at each Dirac point, given by Eq. (\ref{eq:sum}), goes to twice the population at one of the points. If one were able to introduce through optical or electrical means a population difference between the carriers at the two Dirac points so that Eq.(\ref{eq:diff}) is non-zero, there exists the ability to change the direction and magnitude of the anomalous current contribution. 

As the anomalous contribution to the intraband current density goes to zero to first order, the sole first order contribution is due to the first term in Eq. (\ref{eq:intra1}). The first order populations of the conduction and valence bands, $\rho_{nn}^{(1)}(\mathbf{k})$, are proportional to the gradients of their respective Fermi-Dirac distributions, $f_{n}(\mathbf{k},T)$. We can use this relationship to integrate Eq. (\ref{eq:intra1}) by parts.  By changing the integration variable from $\mathbf{k}$ to $\tilde{\epsilon}^{\prime}(\mathbf{k})\equiv\tilde{\epsilon}(\mathbf{k})-t_{\perp},$ we arrive at our final expression for the first order intraband current density:
\begin{equation}
\begin{aligned}\mathbf{J_{i}}^{(1)} & =C(\omega_{p})\int_{0}^{\infty}d\tilde{\epsilon}^{\prime}\left\{ \frac{E_{i}}{E_{b}}\frac{2t_{\perp}^{2}}{(\tilde{\epsilon}^{\prime}+t_{\perp})^{3}}\right.\\
 & +\left.\frac{\left[\left((\tilde{\epsilon}^{\prime}+t_{\perp})(2\tilde{\epsilon}^{\prime}+t_{\perp})+8S^{2}\right)E_{b}^{2}-E_{i}^{2}\right]}{E_{b}^{3}}\frac{\tilde{\epsilon}^{\prime}(\tilde{\epsilon}^{\prime}+2t_{\perp})}{(\tilde{\epsilon}^{\prime}+t_{\perp})^{3}}\right\} \\
 & \times\frac{e^{-\beta E_{b}}+\cosh(\beta E_{b})}{\cosh(\beta E_{F})+\cosh(\beta E_{b})}.
\end{aligned}\label{eq:intra2}
\end{equation}
Here we have defined the variable $E_{i}\equiv E_{0}(\tilde{\epsilon}^{\prime}+t_{\perp})-2S^{2}$ for simplicity. Also, $\beta=k_{B}T$ and $E_{F}$ is the Fermi level of the system. $C(\omega_{p})$ is a time and frequency-dependent coefficient that includes the electric field: $$C(\omega_{p})=\frac{i|e|^{2}\mathbf{E}\left(\omega_{p}\right)e^{-i\omega_{p}t}}{2\hbar^{2}\pi\left(\omega_{p}+i/\tau_{c}\right)},$$ and we take our field to be harmonic, $$\mathbf{E}\left(t\right)=\mathbf{E}\left(\omega_{p}\right)e^{-i\omega_{p}t}.$$  It is possible to show that in the limit that $a\rightarrow0$, Eq. (\ref{eq:intra2}) is identical to the expression for the UBLG case \cite{mcgouran2016nonlinear}.

We follow the same approach for the interband current as we did for the intraband current; calculate the current near each Dirac point, then determine the contribution from both points combined. The first order polarization density near the $\mathbf{K}$-point is given by Eqs. (\ref{eq:4}) and (\ref{eq:connec}):
\begin{equation}
\begin{aligned}\mathbf{P_{e}}^{(1)}(\mathbf{K}) & =4e\sum_{\mathbf{k}}\textrm{Re}\left\{ \left(\frac{E_{0}}{E_{b}}\left(\frac{\tilde{\epsilon}+t_{\perp}}{2\tilde{\epsilon}}\right)\frac{\hat{\mathbf{\theta}}}{k}\right.\right.\\
 & -\left.\left.\frac{iS}{4E_{b}^{2}}\left(\tilde{\epsilon}+2E_{0}\right)\frac{\alpha k}{\tilde{\epsilon}^2}\hat{\mathbf{k}}\right)\rho_{cv}^{(1)}(\mathbf{k})\right\},\label{eq:polar} 
\end{aligned}
\end{equation} 
where we have included the factor of 2 to account for spin degeneracy. The first order matrix element describing the coherence between conduction and valence band is given explicitly by,
\begin{equation}
\rho_{cv}^{\left(1\right)}\left(\mathbf{k}\right)=\frac{e\mathbf{E}\cdot\mathbf{\xi}_{cv}^{b}(\mathbf{k})e^{-i\omega_{p}t}}{\hslash\left(\omega_{cv}(\mathbf{k})-\omega_{p}-i/\tau_{cv}\right)}\left[\rho_{vv}^{\left(0\right)}\left(\mathbf{k}\right)-\rho_{cc}^{(0)}(\mathbf{k})\right].\label{eq:rhocv}
\end{equation}
Here $\omega_{cv}(\mathbf{k})=[E_{c}^{b}(\mathbf{k})-E_{v}^{b}(\mathbf{k})]/\hbar=2E_{b}(\mathbf{k})/\hbar$, and the $\rho_{nn}^{(0)}(\mathbf{k})$ are the zeroth order populations of the conduction and valence bands. From Eq. (\ref{eq:rhocv}), we can see that the first order coherence is proportional to the dot-product of the electric field with the interband connection element: $\mathbf{E}\cdot\mathbf{\xi}_{cv}^{b}(\mathbf{k})$. As such, when converting the sum into an integral via the substitution, $\triangle k\rightarrow0,\;\sum_{k}\rightarrow\frac{\mathnormal{1}}{(2\pi)^{2}}\int d\mathbf{k}$, we must take care when calculating the integral over the angle, $\theta$. To simplify our calculation of the interband polarization density, let us define $\mathbf{\xi}_{vc}^{b}(\mathbf{k})\equiv R(\mathbf{k})\hat{\mathbf{\theta}}-iI(\mathbf{k})\hat{\mathbf{k}}$, where $R(\mathbf{k})$ and $I(\mathbf{k})$ are the amplitudes of the real and imaginary parts of $\mathbf{\xi}_{vc}^{b}(\mathbf{k})$, respectively, 
\begin{equation}
\begin{aligned}R(\mathbf{k}) & =\frac{E_{0}}{E_{b}}\left(\frac{\tilde{\epsilon}+t_{\perp}}{2\tilde{\epsilon}}\right)\frac{1}{k},\\
I(\mathbf{k}) & =\frac{S}{4E_{b}^{2}}\left(\tilde{\epsilon}+2E_{0}\right)\frac{\alpha k}{\tilde{\epsilon}^2}.
\end{aligned}
\end{equation}
Our expression for the first order polarization density around the $\mathbf{K}$-point is then given by,
\begin{equation}
\begin{aligned}\mathbf{P_{e}}^{(1)}(\mathbf{K}) & =\frac{2|e|^{2}}{\hbar}\frac{e^{-i\omega_{p}t}}{(2\pi)^{2}}\int_{0}^{\infty}kdk\left\{ \frac{\left[\rho_{vv}^{\left(0\right)}\left(\mathbf{k}\right)-\rho_{cc}^{(0)}(\mathbf{k})\right]}{\left(\omega_{cv}(\mathbf{k})-\omega_{p}-i/\tau_{cv}\right)}\right.\\
 & \times\left.\int_{0}^{2\pi}d\theta\left(R(\mathbf{k})\hat{\mathbf{\theta}}-iI(\mathbf{k})\hat{\mathbf{k}}\right)\mathbf{E}\cdot\left(R(\mathbf{k})\hat{\mathbf{\theta}}+iI(\mathbf{k})\hat{\mathbf{k}}\right)\right\}\\ 
& +c.c.
\end{aligned}\label{eq:polar2}
\end{equation}
Using the relations $\hat{\mathbf{k}}=cos(\theta)\hat{\mathbf{x}}+sin(\theta)\hat{\mathbf{y}}$, $\hat{\mathbf{\theta}}=-sin(\theta)\hat{\mathbf{x}}+cos(\theta)\hat{\mathbf{y}}$ and $\mathbf{E}=\textrm{E}_x\hat{\mathbf{x}}+\textrm{E}_y\hat{\mathbf{y}}$, and integrating over $\theta$, we obtain
\begin{equation}
\begin{aligned} & \int_{0}^{2\pi}d\theta\left(R(\mathbf{k})\hat{\mathbf{\theta}}-iI(\mathbf{k})\hat{\mathbf{k}}\right)\mathbf{E}\cdot\left(R(\mathbf{k})\hat{\mathbf{\theta}}-iI(\mathbf{k})\hat{\mathbf{k}}\right)\\
 & =\pi\mathbf{E}\left[R(\mathbf{k})^{2}+I(\mathbf{k})^{2}\right]\\
 & -i2\pi R(\mathbf{k})I(\mathbf{k})\left(E_{y}\hat{\mathbf{x}}-E_{x}\hat{\mathbf{y}}\right).
\end{aligned}
\end{equation}
The last term in this integral can be expressed in terms of a cross product with the incident field by noting that $E_{y}\hat{\mathbf{x}}-E_{x}\hat{\mathbf{y}}=\mathbf{E}\times\hat{\mathbf{z}}$. Finally, simplifying once more by employing the relations,
\begin{equation}
\begin{aligned}R(\mathbf{k})^{2}+I(\mathbf{k})^{2} & =\end{aligned}
\mathbf{\xi}_{vc}^{b*}(\mathbf{k})\cdot\mathbf{\xi}_{vc}^{b}(\mathbf{k}),
\end{equation}
\begin{equation}
\begin{aligned}2iR(\mathbf{k})I(\mathbf{k})\hat{\mathbf{z}} & =\mathbf{\xi}_{vc}^{b*}(\mathbf{k})\times\mathbf{\xi}_{vc}^{b}(\mathbf{k}),\end{aligned}
\end{equation}
we can express the first order interband polarization density as,
\begin{equation}
\begin{aligned}\mathbf{P_{e}}^{(1)}(\mathbf{K}) & =D(\omega_{p})\left\{ \mathbf{E}\int_{0}^{\infty}kdk\frac{\mathbf{\xi}_{vc}^{b*}(\mathbf{k})\cdot\mathbf{\xi}_{vc}^{b}(\mathbf{k})}{\left(\omega_{cv}(\mathbf{k})-\omega_{p}-i/\tau_{cv}\right)}N(E_b)\right.\\
 & -\left.\mathbf{E}\times\int_{0}^{\infty}kdk\frac{\mathbf{\xi}_{vc}^{b*}(\mathbf{k})\times\mathbf{\xi}_{vc}^{b}(\mathbf{k})}{\left(\omega_{cv}(\mathbf{k})-\omega_{p}-i/\tau_{cv}\right)}N(E_b)\right\} \\
 & +c.c.
\end{aligned}
\label{eq:313}
\end{equation}
Here we have defined the coefficient $D(\omega_{p})=\frac{|e|^{2}}{2\pi\hbar}e^{-i\omega_{p}t}$. Also, we have defined the zeroth order population difference between the valence and conduction bands as
\begin{equation}
\begin{aligned}N(E_b) & =\rho_{vv}^{\left(0\right)}\left(\mathbf{k}\right)-\rho_{cc}^{(0)}(\mathbf{k})\\
 & =\frac{\sinh(\beta E_{b})}{\cosh(\beta E_{F})+\cosh(\beta E_{b})},
\end{aligned}
\end{equation}

Eq. (\ref{eq:313}) is interesting in the sense that even to first order, the incident field should induce an interband polarization density which has a component that is perpendicular to the direction of the field. There is therefore an \textit{anomalous interband current density} around the $\mathbf{K}$-point. 

Finally, by changing the integration variable from $\mathbf{k}$ to $\tilde{\epsilon}^{\prime}(\mathbf{k}),$ and putting in the full expression for $\mathbf{\xi}_{vc}^{b}(\mathbf{k})$ given in Eq. (\ref{eq:connec}), we obtain for the positive frequency portion of the interband polarization density around the $\mathbf{K}$-point,
\begin{equation}
\begin{aligned}\mathbf{P_{e}}^{(1)}(\mathbf{K}) & =\overleftrightarrow{\chi}(\omega_{p})^{(1)}\mathbf{E}\left(\omega_{p}\right)e^{-i\omega_{p}t},\end{aligned}\label{eq:323}
\end{equation}
where we define the elements of the first order susceptibility matrix $\overleftrightarrow{\chi}(\omega_{p})^{(1)}$ as,
\begin{equation}
\begin{aligned}& \chi^{xx}(\omega_{p})^{(1)} =\\ & \frac{|e|^{2}}{8\pi\hbar}\int_{0}^{\infty}d\tilde{\epsilon}^{\prime}\left\{ \frac{E_{0}^{2}}{E_{b}^{2}}\frac{\left(\tilde{\epsilon}^{\prime}+2t_{\perp}\right)}{\tilde{\epsilon}^{\prime}\left(\tilde{\epsilon}^{\prime}+t_{\perp}\right)\left(\omega_{cv}-\omega_{p}-i/\tau_{cv}\right)}\right.\\
 & +\left.\frac{E_{0}S^{2}\left(2\tilde{\epsilon}^{\prime}+t_{\perp}\right)^{2}\left(\tilde{\epsilon}^{\prime}+2t_{\perp}\right)}{2E_{b}^{4}\left(\tilde{\epsilon}^{\prime}+t_{\perp}\right)^{3}\left(\omega_{cv}-\omega_{p}-i/\tau_{cv}\right)}\right\}N(E_b) \\
\end{aligned}\label{eq:susp1}
\end{equation}
\begin{equation}
\begin{aligned}& \chi^{xy}(\omega_{p})^{(1)} =\\ & \frac{-i|e|^{2}}{8\pi\hbar}\int_{0}^{\infty}d\tilde{\epsilon}^{\prime}\frac{E_{0}S\left(2\tilde{\epsilon}^{\prime}+t_{\perp}\right)\left(\tilde{\epsilon}^{\prime}+2t_{\perp}\right)}{E_{b}^{3}\left(\tilde{\epsilon}^{\prime}+t_{\perp}\right)^{2}\left(\omega_{cv}-\omega_{p}-i/\tau_{cv}\right)}\\
 & \times N(E_b),
\end{aligned}
\end{equation}
\begin{equation}
\begin{array}{cc}
\chi^{yy}(\omega_{p})^{(1)} & =\chi^{xx}(\omega_{p})^{(1)},\\
\chi^{yx}(\omega_{p})^{(1)} & =\chi^{xy*}(\omega_{p})^{(1)}.
\end{array}\label{eq:susp2}
\end{equation}
Here we have suppressed the explicit $\mathbf{k}$-dependencies for simplicity. In the limit that $a\rightarrow0$ (zero bias), we have that $E_{b}\rightarrow E_{0}$, and $S\rightarrow0$. Therefore we recover the polarization density due to the low energy bands in UBLG\cite{mcgouran2016nonlinear}. 

Identical to the cancellation of the anomalous intraband current contribution in Eq. (\ref{eq:intra1}), when we include both the  $\mathbf{K}$ and $\mathbf{K}^{\prime}$-point contributions to the interband current density, the anomalous interband current in Eq. (\ref{eq:313}) goes to zero when the contributions from around both Dirac points are added. This is due to the fact that $I(\mathbf{k})$ has the same sign at each Dirac point but $R(\mathbf{k})$ flips sign, resulting in the zeroth order population difference between $\mathbf{K}$ and $\mathbf{K}^{\prime}$-points (Eq. (\ref{eq:diff})) which is zero. Thus, our full expression for the first order interband polarization should only include the diagonal elements of the susceptibility matrix, $\overleftrightarrow{\chi}(\omega_{p})^{(1)}$, multiplied by two to account for both Dirac points.

From the polarization density, it is simple to obtain an expression for the full first order interband current density including both  $\mathbf{K}$ and $\mathbf{K}^{\prime}$-point contributions. By taking a derivative with respect to time of Eq.(\ref{eq:323}), we have
\begin{equation}
\begin{aligned}\mathbf{J_{e}}^{(1)} & =-2i\omega_{p}\overleftrightarrow{\chi}(\omega_{p})^{(1)}\mathbf{E}\left(\omega_{p}\right)e^{-i\omega_{p}t}\\
 & =\overleftrightarrow{\sigma}(\omega_{p})^{(1)}\mathbf{E}\left(\omega_{p}\right)e^{-i\omega_{p}t},
\end{aligned}\label{eq:320}
\end{equation}
where the first order conductivity matrix is given by $\overleftrightarrow{\sigma}(\omega_{p})^{(1)}=-2i\omega_{p}\overleftrightarrow{\chi}(\omega_{p})^{(1)}$,
and only the diagonal elements of the susceptibility matrix contribute, given by Eqs. (\ref{eq:susp1}) and (\ref{eq:susp2}). The factor of two is due to the contribution from both Dirac points. We can now use Eq. (\ref{eq:320}) to compare to our computer simulation for low field amplitudes. We present the results of our simulations in the next section.

\section{Simulation Results}\label{sec:3}
As was done for the case of UBLG \cite{mcgouran2016nonlinear}, we employ a computer simulation to investigate the nonlinear response of BBLG to THz radiation. One of the major advantages provided by the simulation in the case of BBLG is its ability to examine the dependency of the nonlinear response on the size of the gap between the conduction and valence bands. This band gap is controlled via the external bias. Not only will interband transitions be affected by the gap size (due to the resonance frequency being gap dependent), intraband transitions will also be affected by the distortion of the bands (sombrero feature). The fields transmitted and reflected from the BBLG are calculated as a function of the current densities and the incident field. These fields are then spectrally analyzed to determine their frequency components. A signature of nonlinear behavior is the observation of  high harmonic generation in the spectral composition.

The results of the simulation for low field amplitudes are presented first. We compare these to our first order analytic expressions by showing the agreement between the linear conductivity calculated via both methods.  Finally, we present simulation results which examine the higher order response of BBLG in the presence of an external bias, including both doped and undoped systems.

\subsection{Linear Results}
\begin{figure}[h]
\centering
\includegraphics[width=0.5\textwidth]{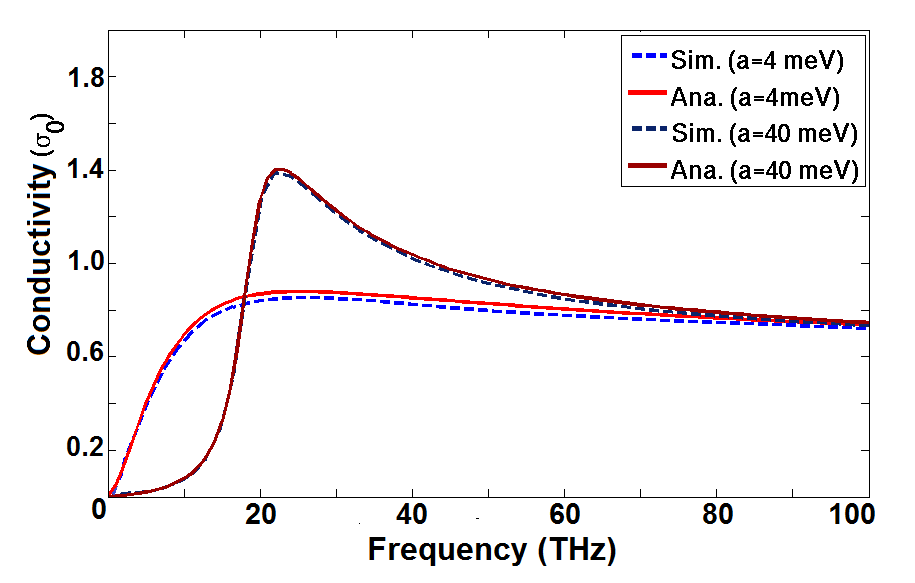}
\caption{Comparison of the real part of the interband conductivity calculated by computer simulation and numerical integration of the closed form expression (Eq.(\ref{eq:320})). Comparison is made at $T$=100 K, $E_F$=0, and $\tau$=50 fs.
  Two different gap sizes are shown, resulting from $a$=4 meV and $a$=40 meV . Conductivity is measured in units of the universal conductivity of UBLG, $\sigma_{0}=e^{2}/2\hbar$\label{fig:5.4}.}
\end{figure}

To begin, we compare the real part of the conductivity due to the interband current density, calculated by both the computer simulation and the closed form expression, Eq. (\ref{eq:320}), for bias values of 4 meV and 40 meV. This comparison is shown in Fig. \ref{fig:5.4} . Since we are comparing an undoped sample, the conductivity associated with intraband transitions is found to be highly suppressed due to the presence of the gap and so we do not include its contribution here. In both cases the scattering time is 50 fs and the temperature is 100 K. As our model for BBLG is an effective two-band model, the features we see in the plot are due solely to transitions between the lower energy conduction and valence bands. One would expect the full four-band model to include features associated with the higher energy transitions - as we see in the case of UBLG \cite{mcgouran2016nonlinear}. In our case, we find that as the incident frequency goes to zero, the conductivity is zero for both bias values. As we increase the incident frequency (measured in THz), we find that the conductivity rises as we approach the resonance of the gap, and ultimately reaches a final value as we increase the frequency further. 

The gap size of 8 meV ($a$=4meV), is approximately equivalent in energy to the photons in a 2 THz pulse. However, at a temperature of 100 K, the thermal energy of the carriers is also approximately 8 meV- equal to the size of the gap. Thus, the interband conductivity at this bias will be affected by thermal populations as well as the THz field driven transitions. We can compare this to the interband conductivity for a gap size of 80 meV ($a$=40 meV). A gap size of this value has a resonance frequency of approximately 20 THz. Due to this gap size being an order of magnitude greater than that for $a$=4 meV, the number of carriers injected by the THz field will be greatly reduced for low photon energies. This leads to a large suppression of the conductivity at lower frequencies compared to the 8 meV gap. Additionally, for $a$=40 meV we see a sharp rise in the conductivity at approximately 20 THz, which is resonant with the size of the gap. This feature is much more sharply peaked than the resonance feature for a=4 meV. This is due to the fact that the conduction and valence bands visibly \textit{flatten} as the external bias is increased (see Fig. \ref{fig:disper}). This flattening leads to a larger number of states being available for the interband transition at this resonant frequency, which increases the spectral weight associated with this transition. This is similar to how the nesting of the conduction bands in UBLG leads to a peak in the conductivity at that resonant frequency (75 THz) \cite{mcgouran2016nonlinear}. Finally, at even higher frequencies, the conductivity for $a$=40 meV reduces to the same value as found for $a$=4 meV. This value is given by $0.5\sigma_{0}$, where $\sigma_{0}$ is the universal conductivity of UBLG, given by $e^{2}/2\hbar$. The reason it does not approach the full universal conductivity at high frequencies is due to the fact our model only considers the transition between the bands $v_2$ and $c_1$. The high-frequency limit of $e^{2}/4\hbar$ is the same as that found for UBLG when we take into account only the low energy transitions\cite{mcgouran2016nonlinear}. If one were to include all the contributions from the higher energy transitions, the interband conductivity should approach $\sigma_{0}$ in the limit of large THz frequencies. 

\subsection{Nonlinear Results}
We now present the results of our full simulations for the nonlinear THz response of BBLG. We have performed simulations for both doped and undoped samples. For the doped case, the incident field amplitude is held fixed and the external bias is set at a number of different values. As we increase the bias, we must increase the Fermi level in order to maintain a consistent carrier density for each case. This allows for a direct evaluation of how THG is affected by the band distortion. For undoped BBLG, we examine the response for a single bias of $a$=2 meV. In this case we keep the bias fixed and adjust the incident field amplitude. From these undoped results we can determine at which field amplitude we expect to observe the largest THG. 

In our simulations for both doped and undoped BBLG, we take the temperature to be 10 K and the scattering time to be 50 fs, which is a conservative value for the temperatures being considered. Our incident field is a sinusoidal Gaussian pulse with central frequency of 1 THz and full width at half maximum (FWHM) of 1 ps. This may be represented mathematically as
\begin{equation}
\mathbf{E}_{i}(t)=\mathbf{E}_{0}e^{-4\textrm{log}(2)\left(\frac{t-t_{0}}{T_{FWHM}}\right)^{2}}\textrm{sin}\left[2\pi f_{0}(t-t_{0})\right],\label{eq:incident}
\end{equation} where $t_{0}$ is the temporal shift and $T_{FWHM}$ is the full width at half maximum of the Gaussian pulse. The central frequency of the pulse is given by $f_{0}$.

For our undoped BBLG simulation, the external bias of $a$=2 meV corresponds to a band gap of 4 meV. The resonant frequency of a band gap this size is approximately 1 THz. Since the incident frequency of our field is on resonance with the band gap, we expect a significant interband current density. However, in the doped case we expect the interband current density to be diminished due to the large Fermi level of the system; the intraband current density will instead be dominant. These two distinct systems allow us to examine both the interband and intraband contributions to the nonlinear response of BBLG. We shall begin by looking at the undoped case.  

\subsubsection{Undoped BBLG: $a$=2 meV}
\begin{figure}[h]
\centering
\includegraphics[width=0.5\textwidth]{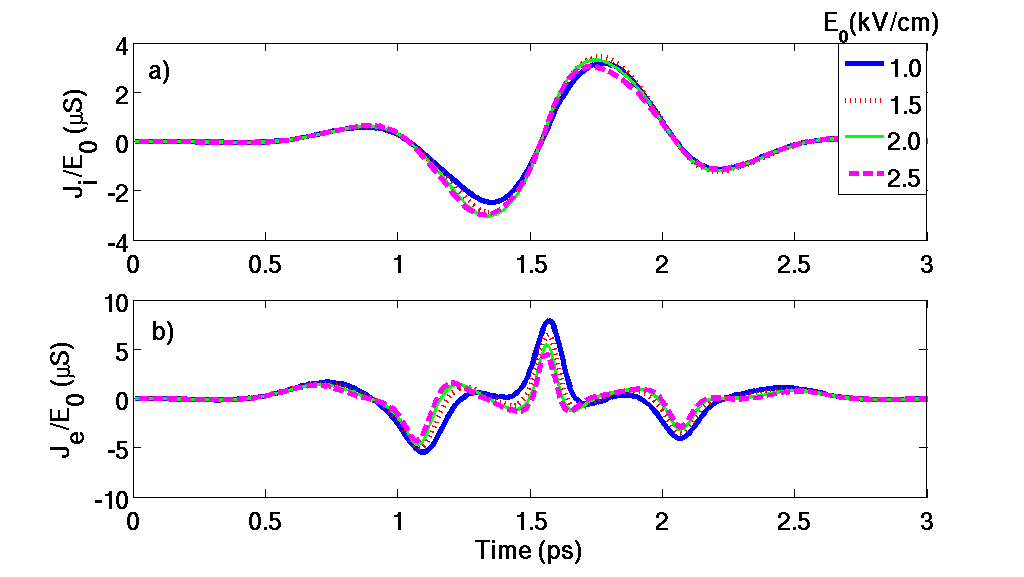}
\caption{Response of BBLG to a incident field of 1 THz at a number of field amplitudes and a bias $a$=2 meV. a) Intraband current density normalized to incident field amplitude. b) Interband current density normalized to incident field amplitude.}\label{fig:5.6}
\end{figure}

We begin by looking at the response of BBLG to incident THz fields with amplitudes ranging from 1.0-2.5 kV/cm, in the presence of an external bias, $a$=2 meV. In order to do so, we examine the interband and intraband current densities at these field amplitudes, followed by the reflected field and the spectral composition of this field (which is dependent on the current densities). In Fig. \ref{fig:5.6} we plot the interband and intraband current densities for four different incident field amplitudes (1.0, 1.5, 2.0, and 2.5 kV/cm). All current densities are normalized to the incident field such that, if the response were linear, these relative currents would be unchanged by an increase in incident field. This procedure allows for a comparison between the current densities at each field amplitude, as well as for the clear identification of any nonlinear behavior. In what follows, we refer to these as \textit{relative current densities}. The relative intraband and interband current densities at these field amplitudes are shown in Figs. \ref{fig:5.6}a and \ref{fig:5.6}b, respectively.

Let us first examine the intraband current density.  We see that as the field amplitude is increased, the intraband current density also increases. This is similar to the behavior of the intraband current density in MLG\cite{al2014high}. This increase arises due to the increase in carrier density from the interband injection of carriers. Since the photons in our 1 THz pulse are essentially resonant on the band gap of 4 meV, this injection of carriers is expected. We present the interband current density for BBLG in Fig. \ref{fig:5.6}b. We clearly see large distortions in the interband current for all of the field amplitudes. Similar to what we observed in our earlier work for the UBLG case\cite{mcgouran2016nonlinear}, we find that as the incident field increases, there is a decrease in the relative interband current density. 

We can also see the effect the external bias has on the ratio of the interband and intraband current densities. For UBLG we found that the ratio of interband to intraband current densities at peak amplitude was approximately $\mathbf{J}_{e}/\mathbf{J}_{i}\simeq0.5$ for an incident field of 1.0 kV/cm \cite{mcgouran2016nonlinear}, i.e. the interband current density is half that of the intraband. However, for BBLG we find that $\mathbf{J}_{e}/\mathbf{J}_{i}\simeq4.0$ for the 1.0 kV/cm field, i.e. the interband current density is approximately four times \textit{larger} than the intraband. Thus, because it opens up a band gap, the application of an external bias allows us to control which current is dominant. Because the interband current contains most of the nonlinearity, we might expect that the application of this bias will increase the overall nonlinear response.

\begin{figure}[h]
\centering
\includegraphics[width=0.5\textwidth]{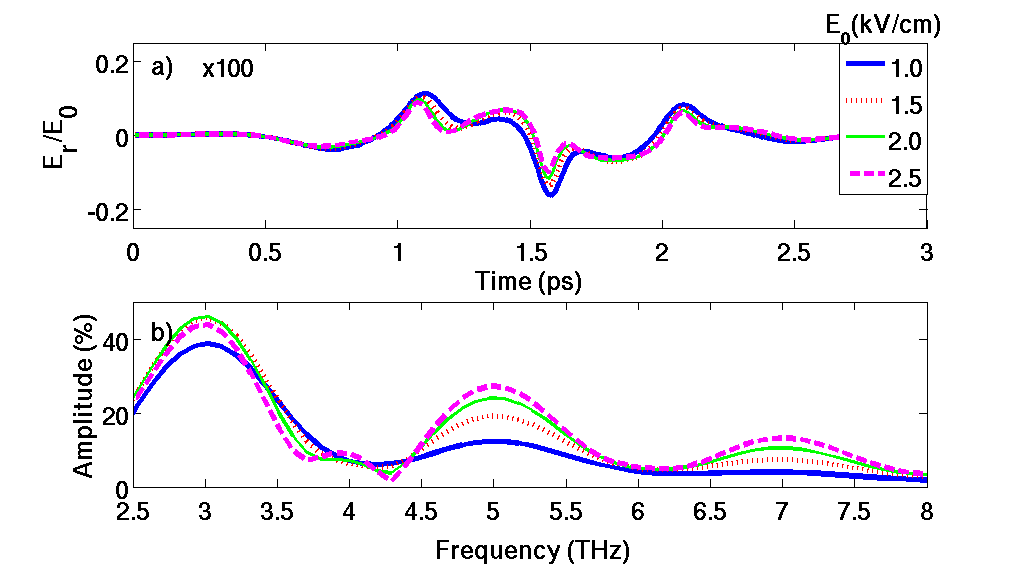}
\caption{Response of BBLG to a incident field of 1 THz at a number of field amplitudes and a bias of $a$=2 meV. a) The reflected field in the temporal domain, normalized to the amplitude of the incident field. Value of $E_r/E_0$ is multiplied by 100 for clarity. b) The amplitude spectra of the reflected signal normalized to the peak at the fundamental frequency of 1 THz.}\label{fig:5.7}
\end{figure}

We can see the effect that the interband and intraband currents have on the nonlinear response of BBLG by looking at the reflected field, as well as its spectral composition. The normalized time-dependent reflected fields for the different field amplitudes (1.0, 1.5, 2.0, and 2.5 kV/cm) are shown in Fig. \ref{fig:5.7}a, and the spectral responses normalized to the peak amplitude of the reflected field at the fundamental frequency (1 THz) are presented in Fig. \ref{fig:5.7}b. 

In Fig. \ref{fig:5.7}a we can see large distortion in the reflected fields for all of the field amplitudes. Furthermore, from the ratio of the current densities given above, it appears that the reflected field for BBLG is dominated by the interband current density, while for UBLG it is dominated by the intraband current density. 

We can see how the differences in the current ratios in biased and unbiased BLG might affect the nonlinear response by looking at the spectral composition of the reflected field for BBLG, in Fig. \ref{fig:5.7}b. We find that at the lowest field amplitude of 1.0 kV/cm, we have a third harmonic signal of approximately 38\% of the reflected spectral peak at the fundamental, which corresponds
to an amplitude of 0.06\% with respect to the fundamental in the incident field. The incident field of 2.0 kV/cm induces the largest third harmonic generation. At this amplitude, we find a maximum in the third harmonic of approximately 45\% the spectral peak at the fundamental (0.07\% of the fundamental in the incident field). For larger field amplitudes the third harmonic is found to decrease again. This value of 45\% of the reflected fundamental is significantly larger than the maximum of 30\% we found in the UBLG case for a 1 THz pulse\cite{mcgouran2016nonlinear}. It is also larger than the value of 32\% found for the third harmonic in MLG under the same conditions\cite{al2014high}. At these higher field amplitudes we also see the presence of a 5th harmonic, which reaches a maximum amplitude of approximately 27\% of the fundamental in the reflected field  (0.03\% of the fundamental in the incident field) for the incident amplitude of 2.5 kV/cm.

The fact that the interband current density plays a dominant role in determining the reflected field in BBLG, may explain the difference between THG in the BBLG and UBLG cases. The nonlinearity is greater in the interband current density, so that when the reflected field is dominated by the interband response, its spectral composition will have a greater percentage of high frequency components. This would explain why we see such a large third harmonic amplitude in the reflected field of BBLG. It also underscores the importance of the interplay between the intraband and interband current densities in producing THG.

We shall next look at the results of simulations for doped BBLG. In this case the response will be due primarily to the intraband current.

\subsubsection{Doped BBLG: $\textrm{E}_{0}=50\,\textrm{kV/cm}$}
\begin{figure}
\centering
\includegraphics[width=0.5\textwidth]{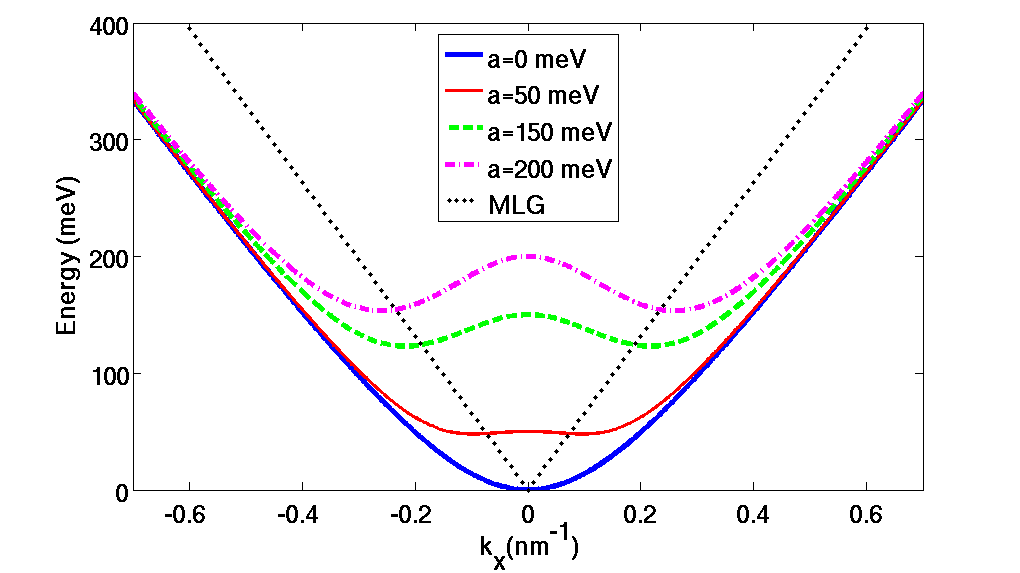}
\caption{Comparison of conduction band, $c_{1}$, of BBLG for bias values of $a=0,\,50,\,150$, and $200\,$meV. The dotted black line shows the band structure of MLG for comparison. At higher bias we see the presence of the 'sombrero' feature.}\label{fig:5.10}
\end{figure}

In this section, we examine the case of doped BBLG. We take the incident field to be the 1 THz pulse given by Eq. (\ref{eq:incident}), with an amplitude of 50 kV/cm. Instead of adjusting the field amplitude as we did for the undoped BBLG simulations, we will adjust the external bias value. We choose four different values of the bias at which to run simulations. These values provide us with a wide range of band structures, allowing us to determine what effect - if any - the curvature of the band has on the nonlinear behavior of doped BBLG. The structure of the conduction band is shown in Fig. \ref{fig:5.10} for the four different biases for which we ran simulations. Also shown in the same figure is the band structure of MLG (dotted black line), which has the characteristic linear dependence on the crystal momentum. As can been seen, the four bias values result in four distinct band structures. For $a$=0 meV, we recover the band structure of UBLG. At $a$=50 meV we see that a gap has been introduced, and there has been some flattening of the band structure. At $a$=150 meV we can clearly start to see the onset of the sombrero feature. For the largest bias of $a$=200 meV, the sombrero feature is well defined and the band gap is now very apparent.  For each of these bias values the charge carrier density is held fixed at a value of $2.0\times10^{12}/cm^{2}$. Of course, for these simulations the size of the gap will not be as important since we are interested in mainly the intraband dynamics. However, the sombrero feature should be of great interest in this analysis.

\begin{figure}
\centering
\includegraphics[width=0.5\textwidth]{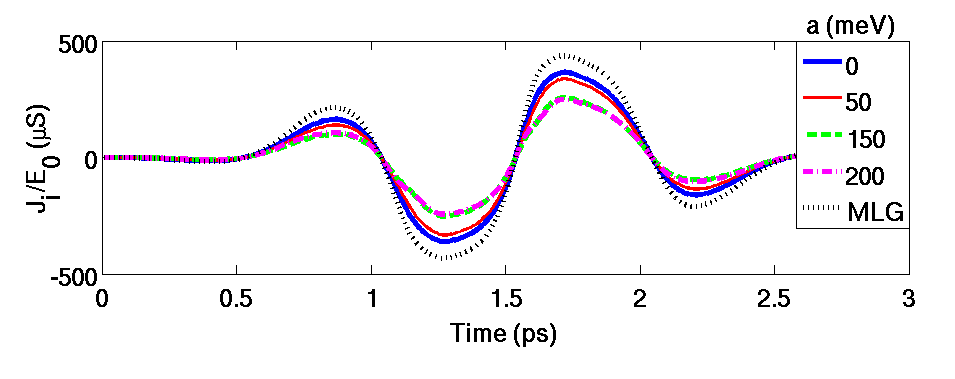}
\caption{Normalized intraband current density of BBLG in response to the 1 THz pulse with amplitude of 50 kV/cm for a number of different bias values, $a=0,\,50,\,150$ and $200\,$meV. Intraband current density normalized to incident field amplitude is shown.The dotted black line shows intraband current density of MLG under identical conditions.}\label{fig:5.10-2}
\end{figure}

Since the interband current density in essentially negligible for the doping level considered, in Fig. \ref{fig:5.10-2} we only plot the relative intraband current densities for the four bias values. We also include the relative intraband current density of MLG for the same carrier concentration for comparison. One thing we notice is that as the bias is increased, the relative intraband current density decreases. From a maximum amplitude of approximately $350\,\mu S$ at zero external bias, to a maximum of approximately $250\,\mu S$ at a bias of 200 meV. The maximum relative current density in MLG at this carrier density is approximately 400$\,\mu S$. It is also quite easy to see the presence of distortion in the intraband current density, for each of the bias values. The shape of this distortion appears to be similar for all biases considered. One might expect the distortions at the higher bias values (150 \& 200 meV) to be of a different nature than those at lower biases. This expectation is based on the sombrero feature being present for large bias values. Certainly the motion of the electrons in the conduction band - which is the basis for our intraband current density - should be affected by this feature. 
\begin{figure}
\centering
\includegraphics[width=0.5\textwidth]{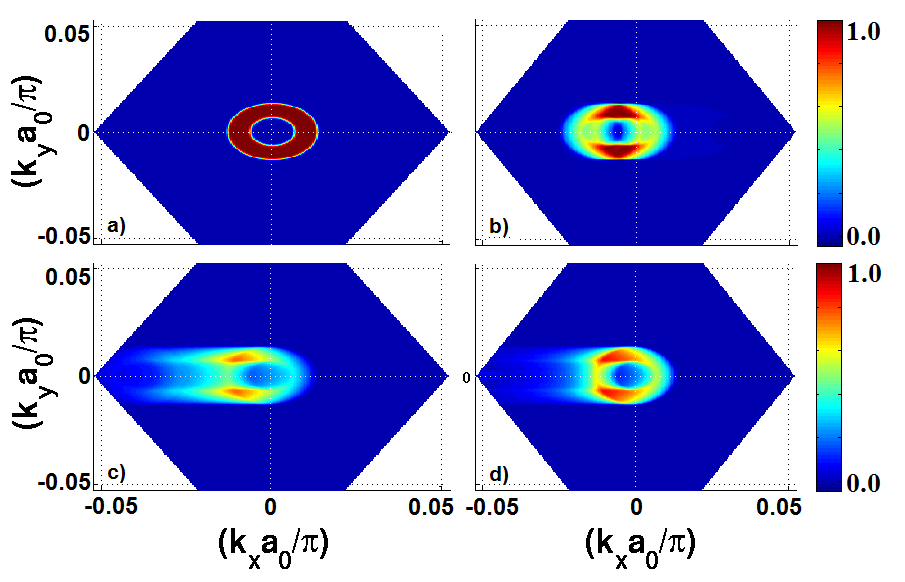}
\caption{Normalized conduction electron density distribution in $k$-space for a bias of $a$=150 meV at (a) the initial thermal conditions, (b) at time t=1.55 ps , (c) at t=1.75 ps, (d) and at t=2.00 ps for a 50 kV/cm incident field. White lines indicate the position of the Dirac point.}\label{fig:5.11}
\end{figure}

\begin{figure}
\centering
\includegraphics[width=0.5\textwidth]{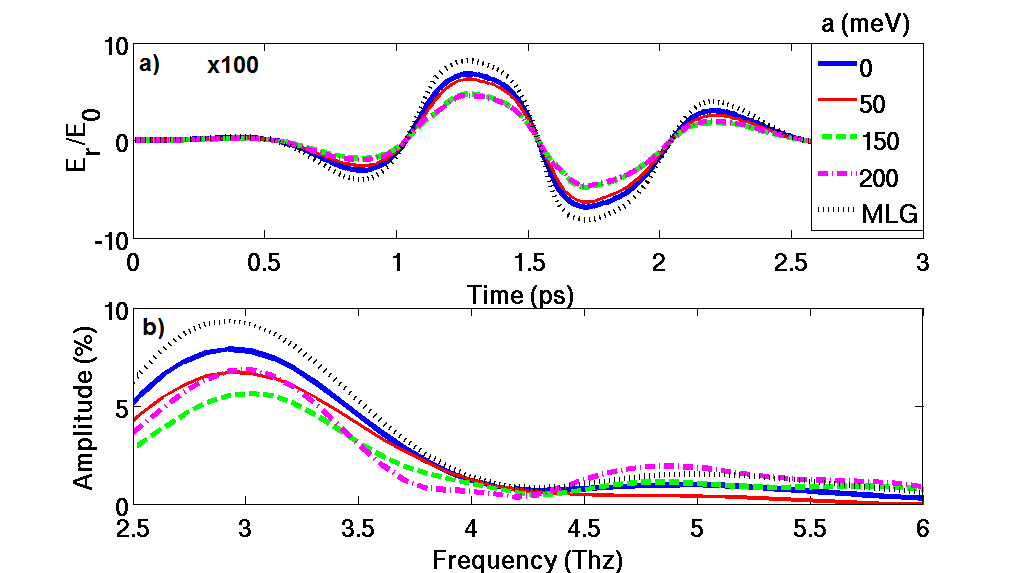}
\caption{Response of BBLG to a incident field of 1 THz at a field amplitude of 50 kV/cm and bias values $a=0,\,50,\,150$ and $200\,$meV. a) The reflected field in the temporal domain, normalized to the amplitude of the incident field. Value of $E_r/E_0$ is multiplied by 100 for clarity. b) The amplitude spectra of the reflected signal normalized to the peak at the fundamental frequency of 1 THz. In both plots, dotted black line shows response of MLG under identical conditions}\label{fig:5.12}
\end{figure}

We can visualize the effect the sombrero feature has on the intraband motion via a plot of the electron density in the conduction band, as shown in Fig. \ref{fig:5.11}. For this simulation we used the bias value of $a$=150 meV, with which we are clearly able to see the sombrero feature present when the carriers are in equilibrium in Fig.  \ref{fig:5.11}a. Here we see that the density of carriers in the conduction band appears as a 'Fermi ring' in $k$-space. We are interested in what happens to the Fermi ring during the interaction with the incident THz field. In Fig.  \ref{fig:5.11}b we see the density at approx t=1.55 ps, after half of the incident pulse has passed. In this instant the distribution of carriers in the conduction band is driven to the left of the Dirac point by the incident field. What is interesting, is that the population has been 'split' in two by the sombrero feature; with half the carriers going above the feature, and half below. Fig.  \ref{fig:5.11}c shows the distribution at t=1.75 ps , at what is essentially its farthest position to the left of the Dirac point. We can see that as the carriers are driven passed the sombrero feature, they begin to merge again on the far side of it. Finally, in Fig.  \ref{fig:5.11}d we see the distribution at t=2.00 ps, when the carriers are moving back towards the Dirac point, and are in the process of forming the ring structure once again. The distribution then settles back into thermal equilibrium (Fig.  \ref{fig:5.11}a) once the pulse leaves the system. Once again, we can determine the effect this 'splitting' of the carrier density has on the nonlinear behavior by looking at the the normalized time-dependent reflected fields for the different bias values, shown in Fig.  \ref{fig:5.12}a, along with the spectral responses normalized to the peak amplitude at the fundamental frequency (1 THz), shown in Fig.  \ref{fig:5.12}b.

Looking at first the reflected field, we see that it is almost 180 degrees out of phase with the intraband current density; which is expected since the interband current is negligible in this case. The maximum amplitude occurs at a bias of $a$=0 meV, i.e when no external bias is present. The value obtained is approximately 3.5 kV/cm for an incident field of 50 kV/cm. We can compare this with the maximum value of 4.0 kV/cm found in MLG at the same incident field amplitude. In terms of the visual distortions in the reflected field, the sombrero feature surprisingly had little effect, other than to decrease the overall maximum amplitude. The shape of the distortions is relatively the same for all bias values (and even MLG). 

We can determine the nonlinear behavior by examining the spectral composition of the reflected fields for these bias values. This is shown in Fig. \ref{fig:5.12}b. It is immediately clear from this plot that the maximum third harmonic, normalized to the peak amplitude at the fundamental frequency (1 THz), is given by the response of MLG; the maximum value is approximately 10\% of the fundamental in the reflected field. This may not be too surprising, as the linear band structure of MLG has been shown to produce highly nonlinear effects \cite{al2014high}\cite{al2015nonperturbative}\cite{al2015optimizing}. More surprising, is the fact that the second largest third harmonic amplitude ($\simeq8\%$ of the reflected fundamental) arises from the case of zero external bias - or the UBLG system. This is surprising in the sense that one would think the sombrero feature - and the splitting effect it has on the intraband motion - would lead to THG that is greater than would be seen in the absence of this feature. This does not seem to be the case for the third harmonic at least, as the bias values of 150 and 200 meV correspond to third harmonic amplitudes of approximately 6\% and 7\% of the fundamental amplitude in the reflected field, respectively. 

However, for all of the bias values considered in the doped case, we find that the ratio of the third harmonic amplitude to the fundamental amplitude in the \textit{incident} field is larger than that found in the undoped case. This is due to the fact that the ratio of the reflected field to the incident field is almost two orders of magnitude larger in the doped case ( Fig. \ref{fig:5.12}a). Specifically, we find that for the four different bias values - $a=0,\,50,\,150$ and $200\,$meV - the third harmonic amplitudes are approximately 0.56\%, 0.40\%, 0.25\%, and 0.30\% of the fundamental amplitude in the incident field, respectively.

\vspace{5mm} 
\section{Summary}\label{sec:4}
We have presented the dynamic equations and results of simulations of the nonlinear response of undoped and doped BBLG at THz frequencies. The central goal of this work was to determine the role that the external bias plays in the nonlinear response. To model the response, a theoretical model was developed based on the dynamic equations of density matrix elements within the basis of an effective Hamiltonian. This allowed for the calculation of the eigenvalues and eigenvectors of BBLG, as well as the interband connection elements, Berry connections and Berry curvatures of the band structure. Expressions for interband and intraband current densities were also derived.  

Solutions to the density matrix dynamic equations were determined through the use of simulation. These solutions were then applied to the study of high harmonic generation in undoped and doped BBLG for a number of external bias values. The undoped system allowed us to investigate the interplay between interband and intraband dynamics, and what effect it has on THG. The doped system allowed us to determine whether or not the unique band structure of BBLG - specifically the sombrero feature - has any influence on the nonlinear behavior. 

Our results show that for undoped BBLG, the largest third harmonic amplitude for a 1 THz single-cycle pulse was found to be 45\% of the peak fundamental amplitude in the reflected field (0.07\% of the fundamental in the incident field) for an external bias of 2 meV. We also found that the ratio of the interband and intraband current densities is affected by the value of the external bias, and that this ratio may play an important role in THG. Finally, we showed that for a doped system, the amplitude of the third harmonic reaches a maximum of 8\% of the fundamental in the reflected field (0.56\% of the fundamental in the incident field) for zero bias, and decreases as we increase the external bias.

To experimentally observe the high harmonics we predict for BBLG, one must consider the dynamic range of the THz spectrometers - defined as the ratio of the frequency dependent signal strength to the detected noise floor\cite{jepsen2005dynamic}. For a 1 THz  incident field of 1.0 kV/cm, we find the peak amplitude of the reflected field from undoped BBLG to be approximately 1.7 V/cm (55 dB less than the incident field). Thus, a detection technique that allows for a dynamic range larger than 55 dB is required for the measurement of the reflected signal. This can be achieved, as a very high dynamic range of 90 dB has recently been reported\cite{vieweg2014terahertz}. 

\bibliographystyle{apsrev4-1}

\bibliography{UCThesisBibliography}

\end{document}